# Local perfect chirality at reflection-zeros away from exceptional points in optical whispering gallery microcavity

Junda Zhu,[1,2] Haitao Liu,[3,4,*] Fang Bo,[2,†] Can Tao,[2,§] Guoquan Zhang,[2] and Jingjun Xu[2]
[1]*College of Physics and Materials Science, Tianjin Normal University, Tianjin 300387, China*
[2]*The MOE Key Laboratory of Weak Light Nonlinear Photonics, TEDA Institute of Applied Physics and School of Physics, Nankai University, Tianjin 300457, China*
[3]*Institute of Modern Optics, College of Electronic Information and Optical Engineering, Nankai University, Tianjin 300350, China*
[4]*Tianjin Key Laboratory of Micro-scale Optical Information Science and Technology, Tianjin 300350, China*

Recently, a local and imperfect chirality of the resonant eigenmode at the exceptional point (EP) has been reported in the optical whispering gallery microcavity system perturbed by two strong nanoscatterers [Phys. Rev. A **108**, L041501 (2023)]. Here, we discover a local perfect chirality of the resonant eigenmode away from the EP in the parameter space of the strongly perturbed microcavity system. By considering the multiple scattering process of the azimuthally propagating modes (APMs) at the nanoscatterers with a first-principles-based model, the local perfect chirality is predicted to result from the unidirectional reflectionlessness, i.e., the reflection-zero (*R*-zero) of the APMs at the two nanoscatterers. Numerical results and model predictions consistently show that the structural parameters of the *R*-zero typically deviate from those of the EP, which means that the pair of split resonant eigenmodes at the *R*-zero have different complex resonance frequencies and electromagnetic fields. In general, only one of the pair of split eigenmodes exhibits a local perfect chirality within the local azimuthal range divided by the two nanoscatterers. With the decrease of the two nanoscatterers' sizes or their relative azimuthal angle, the *R*-zero tends to coincide with the EP.

## I. INTRODUCTION

Unidirectional reflectionlessness [1] is an intriguing phenomenon in non-Hermitian optical system, which has been widely studied in multiple-channels scattering system, such as parity-time symmetric Bragg scatterer systems [2-4], metasurface systems [5,6] and coupled-resonator systems [7], and has been used to realize unidirectional invisibility [2-4], coherent perfect absorber [8-11] and reflectionless scattering mode [12,13]. Generally, by tuning the structural parameters of the non-Hermitian scattering system, the unidirectional reflectionlessness could occur at discrete and complex-valued incident frequencies for certain input channels, which are termed as *reflection-zeros* (*R*-zeros) [5,6,12-18].

In the optical whispering gallery microcavity perturbed by asymmetric nanoscatterers (e.g., nanoparticles [19], nanodefects [20] or nanoholes [21]), the weight between the clockwise (CW) and counterclockwise (CCW) travelling waves in forming the resonant eigenmode of the perturbed microcavity is unbalanced, which is termed as the chirality of electromagnetic field of the resonant eigenmode [22-26] and has important applications such as photon blockade [19], suppression of backscattering [20] and chiral lasing [24]. By modeling the travelling waves with a global basis of whispering gallery modes (WGMs) defined as the eigenmodes of unperturbed microcavity, the chirality is attributed to the asymmetric coupling between the CW and CCW WGMs at the nanoscatterers [23]. By modeling the travelling waves with a local basis of azimuthally propagating modes (APMs) defined as the waveguide modes supported by a mapped azimuthally-invariant waveguide [27], the chirality can be understood as resulting from the asymmetric reflection of the APMs at the nanoscatterers [28]. Specially, by precisely tuning two *weak* nanoscatterers, a fully asymmetric reflection, namely, a unidirectional reflectionlessness (i.e., *R*-zero) of the APMs at the nanoscatterers, can be realized at the resonant exceptional point (EP) of the system [28]. The complex resonance frequencies degenerate and the corresponding pair of resonant eigenmodes coalesce at the EP [29,30]. Consequently, the coalesced resonant eigenmode at the EP is predicted to be formed by a pure CW or CCW travelling wave of WGM [23] or APM [28], which corresponds to a perfect global chirality over the whole azimuthal range from 0 to $2\pi$.

Recently, it is reported that for the microcavity perturbed by two *strong* nanoscatterers of nanoholes, the chirality of the resonant eigenmode at the EP is generally local (i.e., the chiralities within the two azimuthal ranges divided by the two scatterers are different) and imperfect due to the frequency-dependent scattering of the APMs at the nanoscatterers [31], where the local basis of APMs is used to describe the local chirality for which the global basis of WGMs for describing the global chirality is not applicable. With the decease of the size of the scatterers or the relative azimuthal angle between the two scatterers, the chirality shows a *tendency* to be globally or locally perfect, respectively. However, an exactly perfect local chirality of the resonant eigenmode along with its relation with the EP has not been studied yet in the system.

In this paper, we reveal an *exactly perfect* local chirality of the resonant eigenmode at the *R*-zero that is generally away from the EP in the microcavity system perturbed by two *strong* nanoscatterers of nanoholes. By mapping the perturbed microcavity into the cylindrical coordinate system, the multiple scattering problem of the CW and CCW APMs at the nanoscatterers can be treated as a two-ports scattering problem in a mapped azimuthally-invariant waveguide. With a first-principles-based model, an exactly perfect local chirality is predicted to be achieved at the *R*-zero of the system, at which a unidirectional reflectionlessness of the APMs occurs at the nanoscatterers. In general, the structural parameters of the *R*-zero are different from those of the EP. At the *R*-zero, the system supports a pair of split resonant eigenmodes with different complex resonance frequencies and electromagnetic fields. Only one of the split eigenmodes exhibits a locally perfect chirality. Different from the local chirality at the EP, where the stronger local chirality only exists in the larger azimuthal range divided by the two nanoscatterers [31], the locally perfect chirality at the *R*-zero could exist in either the larger or the smaller azimuthal range. The *R*-zero will tend to coincide with the EP under the condition of weak scatterers or small relative azimuthal angle between the two scatterers.

This article is organized as follows. In Sec. II, we introduce the system, the definition of local chirality and the theoretical model for the *R*-zero. In Sec. III, we provide the numerical results of the local chirality around the resonant EP and *R*-zero, and the impact of the sizes of the two nanoscatterers and their relative azimuthal angle on the local chirality at the *R*-zero. We summarize our results in Sec. IV.

## II. THEORETICAL MODEL

### A. The system and the definition of the local chirality

Here we consider a two-dimensional cylindrical microcavity of radius $R_0$ perturbed by two nanoholes near the boundary of the microcavity [28,31,32]. The present system is chosen similar to that in a recent work on the local chirality at the resonant EPs [31], so that we can directly compare the local chiralities and the resonance frequencies of the eigenmodes at the *R*-zeros with those at the EPs. As sketched in Fig. 1(a), $r_j$, $\theta_j$ and $d_j$ ($j=1, 2$) denote the radial lengths, azimuthal ranges and distances to the cavity boundary for the two sectorial-shaped nanoholes, respectively.

To describe the local chirality of the resonant eigenmodes in different azimuthal ranges divided by the two nanoholes, we choose a pair of CCW and CW APMs ($\Psi_{cc}$, $\Psi_c$) as the local basis, where $\Psi_{cc}=\Psi_{cc,r}(r)\exp(ik_0n_{eff}R_0\phi)$ and $\Psi_c=\Psi_{c,r}(r)\exp[ik_0n_{eff}R_0(2\pi-\phi)]$ denote the electric and magnetic field vectors $\Psi=[\mathbf{E},\mathbf{H}]$ of the normalized CCW and CW APMs, respectively. Because the APMs can be rigorously defined as the waveguide modes supported by a mapped straight waveguide along $\phi$-direction as explained hereafter [27], the APMs can be used to describe the local (i.e., $\phi$-dependent) electromagnetic field of the eigenmode [31], with their complex effective index $n_{eff}$ and electromagnetic field all depending on the frequency $\omega$.

As sketched in Fig. 1(a), $a_{cc(c)}$ and $b_{cc(c)}$ denote the complex-amplitude coefficients of the CCW (CW) APMs within the azimuthal ranges from $\beta$ to $2\pi$ and from 0 to $\beta$, respectively. The local field chiralities of the resonant eigenmodes are defined as $\alpha_{\beta\sim 2\pi}=(|a_{cc}/v|^2-|a_c|^2)/(|a_{cc}/v|^2+|a_c|^2)$ and $\alpha_{0\sim\beta}=(|b_{cc}|^2-|b_c/w|^2)/(|b_{cc}|^2+|b_c/w|^2)$ from $\beta$ to $2\pi$ and from 0 to $\beta$, respectively [31]. Here $v=\exp(ik_0n_{eff}R_0\beta)$ and $w=\exp[ik_0n_{eff}R_0(2\pi-\beta)]$ are the phase-shift factors of the APM traveling azimuthally from 0 to $\beta$ and from $\beta$ to $2\pi$, respectively.

### B. Theoretical model for the *R*-zeros

To provide an intuitive and quantitative description of the scattering problem of the APMs at the scatterers, we will derive a theoretical model with a scattering-matrix method. Under the cylindrical coordinate system $(r,\phi,z)$ with $x=r\cos\phi$ and $y=r\sin\phi$, the microcavity perturbed by sectorial-shaped nanoholes is mapped into a straight waveguide with periodically embedded rectangular nanoholes along the $\phi$ direction as sketched in Fig. 1(b). Then the APMs can be rigorously defined as the waveguide modes supported by the mapped $\phi$-invariant waveguide [27]. Due to the periodic boundary condition along the $\phi$ direction, the APM coefficients ($a_{cc}$, $a_c$, $b_{cc}$ and $b_c$) have a period of $2\pi$. The two nanoholes along with the azimuthal range from 0 to $\beta$ ($\beta$ to $2\pi$) are defined as the effective (complementary effective) scatterer, as illustrated by the blue (red) region in Fig. 1(b).

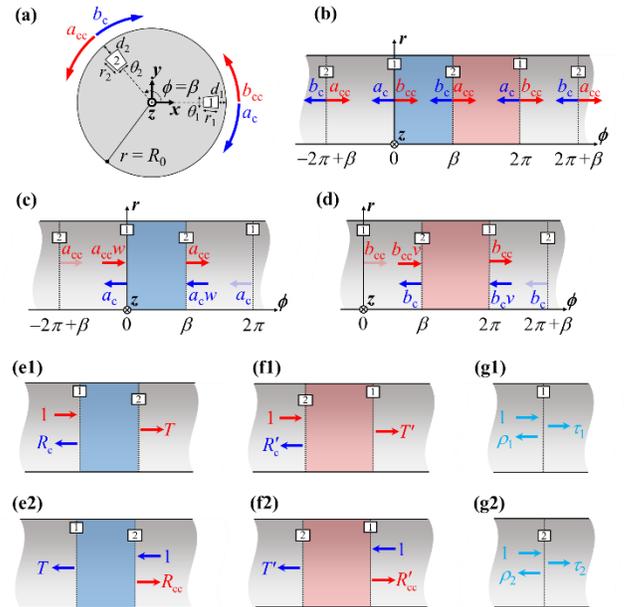

FIG. 1. (a) Schematic of the microcavity perturbed by two sectorial-shaped nanoholes under the rectangular Cartesian coordinates $(x,y,z)$. (b) Mapping the system sketched in (a) into a $\phi$-invariant waveguide with periodically embedded rectangular nanoholes under the cylindrical coordinate system $(r,\phi,z)$. The blue and red regions denote the effective scatterer and complementary effective scatterer in one azimuthal period $(0,2\pi)$, respectively. (c)-(d) Scattering problems of the APMs at the effective scatterer and complementary effective scatterer, respectively. (e1)-(e2) Definition of the effective scattering coefficients $(R_c, R_{cc}, T)$ of APMs at the effective scatterer. (f1)-(f2) Definition of the complementary effective scattering coefficients $(R'_c, R'_{cc}, T')$ of APMs at the complementary effective scatterer. (g1)-(g2) Definition of the reflection ($\rho_j$) and transmission ($\tau_j$) coefficients of APMs at nanohole 1 and 2, respectively.

We first focus on the scattering problem of the APMs at the effective scatterer. As shown in Fig. 1(c), the scattering process at the effective scatterer can be treated as a two-ports scattering problem under the cylindrical coordinate system. The coefficients $\mathbf{a}_{in}$ and $\mathbf{a}_{out}$ for the incoming and outgoing APMs at the effective scatterer are related through a frequency-dependent scattering matrix $\mathbf{S}(\omega)$,

$$\mathbf{a}_{out} = \mathbf{S}(\omega)\mathbf{a}_{in}. \quad (1)$$

Here $\mathbf{a}_{out}=(a_{cc},a_c)^T$ according to the definition of $a_{cc}$ and $a_c$ as shown by the outgoing arrows at the effective scatterer in Fig. 1(c). The coefficients of incoming APMs at $-2\pi+\beta$ and $2\pi$ (shown by the dashed arrows) are respectively given by $a_{cc}$ and $a_c$ due to the periodicity of the APM fields. Thus, the coefficients of incoming APMs at the effective scatterer are given by $\mathbf{a}_{in}=(a_{cc}w,a_cw)^T=w\mathbf{a}_{out}$ (shown by the incoming solid arrows) with an additional phase-shift factor $w=\exp[ik_0 n_{eff}R_0(2\pi-\beta)]$ for the APM travelling over $2\pi-\beta$. The scattering matrix $\mathbf{S}(\omega)$ can be written as,

$$\mathbf{S}(\omega) = \begin{bmatrix} T(\omega) & R_{cc}(\omega) \\ R_c(\omega) & T(\omega) \end{bmatrix}, \quad (2)$$

where $R_{cc}$ ($R_c$) denotes the effective reflection coefficient of the APM for a CW (CCW) incident APM, and $T$ denotes the effective transmission coefficient as defined in Figs. 1(e1)-(e2). Note that $T$ for a CCW incident APM [Fig. 1(e1)] is equal to $T$ for a CW incident APM [Fig. 1(e2)] according to the reciprocity theorem of waveguide modes [33]. With a Fabry-Pérot-like model, the effective scattering coefficients $R_{cc}$, $R_c$ and $T$ can be further expressed as [28],

$$R_{c(cc)} = \rho_{1(2)} + \frac{v^2 \rho_{2(1)} \tau_{1(2)}^2}{1-v^2\rho_1\rho_2}, \quad T = \frac{v\tau_1\tau_2}{1-v^2\rho_1\rho_2}, \quad (3)$$

where $\rho_j$ and $\tau_j$ ($j=1, 2$) denote the scattering coefficients of the APM at each nanohole as defined in Figs. 1(g1)-(g2). The $\rho_j$ and $\tau_j$ along with the electromagnetic field and complex effective index $n_{eff}$ of the APMs at a given frequency $\omega$ are rigorously calculated with a full-wave aperiodic Fourier modal method (a-FMM) under the cylindrical coordinate system [27,34]. This rigorous calculation of model parameters based on the first principle of Maxwell's equations without any fitting process ensures that the model has a solid electromagnetic foundation and thus can provide quantitative predictions. $\rho_j$ and $\tau_j$ weakly depend on $\omega$ [27], but the phase-shift factor $v=\exp(ik_0 n_{eff}R_0\beta)$ typically depends on $\omega$ via $k_0=\omega/c$ (except for some special cases, such as weak scatterer or small $\beta$ [31]), so that the scattering matrix $\mathbf{S}(\omega)$ typically depends on $\omega$. Note that Eq. (1) is equivalent to the coupled-APMs equations built in Refs. [28,31].

Substituting $\mathbf{a}_{in}=w\mathbf{a}_{out}$ into Eq. (1), one can obtain,

$$\left[\mathbf{S}(\omega) - \frac{1}{w(\omega)}\mathbf{I}\right]\mathbf{a}_{out} = 0, \quad (4)$$

which forms a nonlinear eigenvalue problem with $\omega$ being the eigenvalue. By setting the determinant of the coefficient matrix of Eq. (4) to be zero, i.e., det$[\mathbf{S}(\omega)-1/w(\omega)\mathbf{I}]=0$, one can obtain,

$$w(\omega) = \frac{1}{T(\omega) \pm \sqrt{R_{cc}(\omega)R_c(\omega)}}, \quad (5)$$

where $\pm$ denotes one of the two single-valued branches of the square root function. The complex resonance frequencies $\omega_{1,2}$ of the two split eigenmodes can be respectively determined by solving the two transcendental equations (5) with numerical iterative method, e.g., the iterative linear-interpolation method [28]. Substituting the solved eigenvalues $\omega=\omega_1$ or $\omega_2$ into Eq. (4), the associated eigenstates $\mathbf{a}_{out}$ can be expressed as [assuming $R_c(\omega_{1,2})\neq 0$],

$$\begin{bmatrix} a_{cc}(\omega_{1,2}) \\ a_c(\omega_{1,2}) \end{bmatrix} = \begin{bmatrix} \pm\sqrt{R_{cc}(\omega_{1,2})/R_c(\omega_{1,2})} \\ 1 \end{bmatrix}. \quad (6)$$

Then the local chirality $\alpha_{\beta\sim 2\pi}$ in the azimuthal range from $\beta$ to $2\pi$ for the pair of split eigenmodes can be written as

$$\alpha_{\beta\sim 2\pi}(\omega_{1,2}) = \frac{|R_{cc}(\omega_{1,2})|-|R_c(\omega_{1,2})||v(\omega_{1,2})|^2}{|R_{cc}(\omega_{1,2})|+|R_c(\omega_{1,2})||v(\omega_{1,2})|^2}. \quad (7)$$

When the reflectionlessness of the APM occurs at a certain complex resonance frequency $\omega_R$, there is $R_c(\omega_R)=0$ or $R_{cc}(\omega_R)=0$, which corresponds to the $R$-zeros of the scattering problem [12]. Generally, the $R$-zeros $R_c(\omega_R)=0$ and $R_{cc}(\omega_R)=0$ would not arise simultaneously due to the lack of symmetry of the effective scatterer, so the reflectionlessness of the APM at the effective scatterer is unidirectional. According to Eqs. (6) and (7), the APM coefficients $[a_{cc}(\omega_R),a_c(\omega_R)]^T$ of the resonant eigenmode at the $R$-zeros $R_c(\omega_R)=0$ or $R_{cc}(\omega_R)=0$ will reduce to $(1,0)^T$ or $(0,1)^T$, which then results in a locally perfect chirality $\alpha_{\beta\sim 2\pi}=1$ or $-1$, respectively.

The $R$-zeros of the scattering problem at the effective scatterer can be determined by the following equations,

$$R_c(\omega_R) = 0 \text{ or } R_{cc}(\omega_R) = 0, \quad (8a)$$
$$w(\omega_R)T(\omega_R) = 1, \quad (8b)$$

where Eq. (8b) is obtained by substituting Eq. (8a) into Eq. (5). Equation (8b) can be understood intuitively. When the reflectionlessness of the APM occurs at the effective scatterer, $R_c=0$ for instance, the outgoing CW APM (with coefficient $a_c$) only results from the transmission ($T$) of the incoming CW APM (with coefficient $a_c w$) in the absence of the reflection ($R_c$) of the incoming CCW APM, i.e., $a_c=a_c wT$, as shown in Figs. 1(c) and (e1)-(e2). The structural parameters ($r_1$ and $\beta$ for instance) and the complex resonant frequency $\omega_R$ at the $R$-zeros $R_c(\omega_R)=0$ or $R_{cc}(\omega_R)=0$ can be obtained by solving the complex-valued transcendental equations (8) with numerical iterative method [28].

Similarly, by considering the scattering of the APMs at the complementary effective scatterer as shown in Fig. 1(d), one can obtain the nonlinear eigenvalue problem

$$\left[\mathbf{S}'(\omega) - \frac{1}{v(\omega)}\mathbf{I}\right]\mathbf{b}_{out} = 0, \quad (9)$$

where $\mathbf{b}_{out}=(b_{cc},b_c)^T$ and

$$\mathbf{S}'(\omega) = \begin{bmatrix} T'(\omega) & R'_{cc}(\omega) \\ R'_c(\omega) & T'(\omega) \end{bmatrix}. \quad (10)$$

Here $R'_{cc}$, $R'_c$ and $T'$ denote the complementary effective scattering coefficients of the APM at the complementary effective scatterer as defined in Figs. 1(f1)-(f2), and can be expressed as [28],

$$R'_{c(cc)} = \rho_{2(1)} + \frac{w^2 \rho_{1(2)} \tau^2_{2(1)}}{1-w^2 \rho_1 \rho_2}, \quad T' = \frac{w\tau_1\tau_2}{1-w^2\rho_1\rho_2}, \quad (11)$$

By solving Eq. (9), one can obtain

$$v(\omega) = \frac{1}{T'(\omega) \pm \sqrt{R'_{cc}(\omega)R'_c(\omega)}}, \quad (12)$$

and

$$\begin{bmatrix} b_{cc}(\omega_{1,2}) \\ b_c(\omega_{1,2}) \end{bmatrix} = \begin{bmatrix} \pm\sqrt{R'_{cc}(\omega_{1,2})/R'_c(\omega_{1,2})} \\ 1 \end{bmatrix}. \quad (13)$$

Then the local chirality $\alpha_{0\sim\beta}$ in the azimuthal range from 0 to $\beta$ for the pair of split eigenmodes can be written as

$$\alpha_{0\sim\beta}(\omega_{1,2}) = \frac{|w(\omega_{1,2})|^2 |R'_{cc}(\omega_{1,2})| - |R'_c(\omega_{1,2})|}{|w(\omega_{1,2})|^2 |R'_{cc}(\omega_{1,2})| + |R'_c(\omega_{1,2})|}. \quad (14)$$

According to Eq. (14), the resonant eigenmodes at the $R$-zero $R'_c(\omega_R) = 0$ or $R'_{cc}(\omega_R) = 0$ have a locally perfect chirality $\alpha_{0\sim\beta}=1$ or $-1$, respectively. The $R$-zeros of the scattering problem at the complementary effective scatterer can be determined by the equations,

$$R'_c(\omega_R) = 0 \text{ or } R'_{cc}(\omega_R) = 0, \quad (15a)$$

$$v(\omega_R)T'(\omega_R) = 1. \quad (15b)$$

Note that Eqs. (8) or (15) only determine one of the split eigenmodes at the $R$-zero with a locally perfect chirality $|\alpha_{\beta\sim 2\pi}|=1$ or $|\alpha_{0\sim\beta}|=1$, respectively. The other eigenmode with a different complex resonant frequency, which generally does not satisfy the unidirectional reflectionless condition (8a) or (15a), should be determined by solving the transcendental equation (5) or (12). Exceptions are the resonant EPs with a small relative azimuthal angle $\beta$ between the two scatterers, at which the pair of split eigenmodes coalesce into one eigenmode with an approximately perfect local chirality in the larger azimuthal range divided by the two scatterers [31].

Besides, the $R$-zeros $R_{cc(c)}(\omega_R) = 0$ at the effective scatterer and $R'_{cc(c)}(\omega_R) = 0$ at the complementary effective scatterer typically would not arise simultaneously. Exceptions are the resonant EPs with weak scatterers, at which the pair of split eigenmodes coalesce into one eigenmode with an approximately perfect global chirality over the whole azimuthal range $0\sim 2\pi$ [31].

## III. NUMERICAL RESULTS
### A. Local chirality around resonant EP and $R$-zero

For the numerical calculation, the resonant eigenmodes are solved with the full-wave finite element method (FEM) performed by COMSOL MULTIPHYSICS software. The electromagnetic field and complex effective index $n_{eff}$ of the APM at the complex resonance frequency of eigenmode are solved with the full-wave a-FMM under the cylindrical coordinate system [27,34]. Then the local chirality $\alpha_{\beta\sim 2\pi}$ or $\alpha_{0\sim\beta}$ can be calculated by extracting the APM coefficients $a_{cc}$, $a_c$, $b_{cc}$ and $b_c$ from the electromagnetic field of the eigenmodes with the mode-orthogonality theorem [33].

We first calculate the local chirality of the resonant eigenmodes around an EP already shown in Ref. [31]. The structural parameters of the system are set to be $R_0$=1.6 μm, $d_1$=0.04 μm, $d_2$=0.048 μm, $r_2$=0.27204 μm, $\theta_j$=2arcsin[$r_j$/(4$R_0$)] ($j$ = 1, 2). The refractive indices of the microcavity, the nanoholes and the surrounding medium are set to be 2, 1 and 1 (air), respectively. For solving the EP, two structural parameters $r_1$ and $\beta$ are scanned and their values at the EP are listed in TABLE I.

TABLE I. Numerical results of structural parameters $r_1$ and $\beta$, dimensionless complex resonance frequency shifts $\Omega_{1(2)}-\Omega_0$ and local chiralities $\alpha_{\beta\sim 2\pi,1(2)}$ and $\alpha_{0\sim\beta,1(2)}$ (with an additional subscript 1 or 2 for eigenmode 1 or 2) at the EP (already obtained in Ref. [31]) and $R$-zeros.

|  | EP[31] | $R$-zero $R_{cc}(\omega_R) = 0$ | $R$-zero $R'_{cc}(\omega_R) = 0$ |
|---|---|---|---|
| $r_1$ (μm) | 0.2780960 | 0.2778880 | 0.2767260 |
| $\beta$ (rad) | 1.6838536 | 1.6838500 | 1.6838109 |
| $\Omega_1-\Omega_0$ | 0.0794348 | 0.0794415 | 0.0792107 |
|  | -0.0075841i | -0.0072938i | -0.0068357i |
| $\Omega_2-\Omega_0$ | 0.0794195 | 0.0793080 | 0.0789619 |
|  | -0.0075777i | -0.0078534i | -0.0082314i |
| $\alpha_{\beta\sim 2\pi,1}$ | -0.9831389 | -0.9348146 | -0.8068930 |
| $\alpha_{0\sim\beta,1}$ | -0.8693274 | -0.9459416 | *-0.9999995* |
| $\alpha_{\beta\sim 2\pi,2}$ | -0.9828323 | *-0.9999996* | -0.9547775 |
| $\alpha_{0\sim\beta,2}$ | -0.8703013 | -0.7738531 | -0.6011465 |

Figure 2(a) shows the numerical results of the local chiralities $\alpha_{\beta\sim 2\pi}$ (solid curves) and $\alpha_{0\sim\beta}$ (dashed curves) for the pair of eigenmodes plotted as functions of $r_1$ around the EP (vertical green dotted line) with other parameters fixed. The dimensionless complex resonance frequency $\Omega=\omega R_0/c$ of the eigenmodes are plotted in Fig. 2(b) as a function of $r_1$.

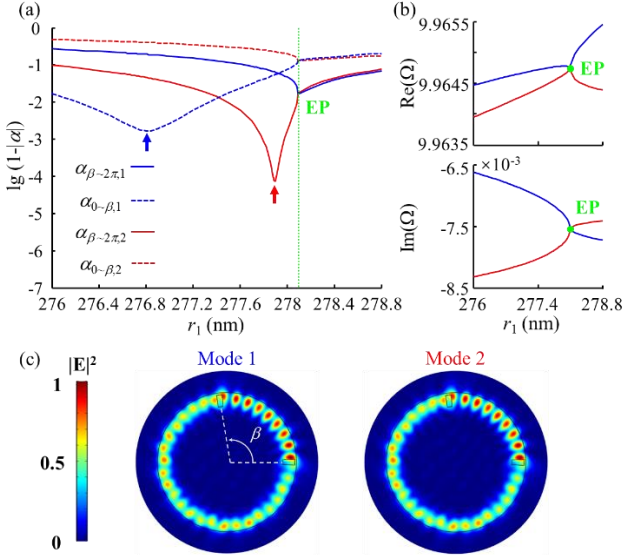

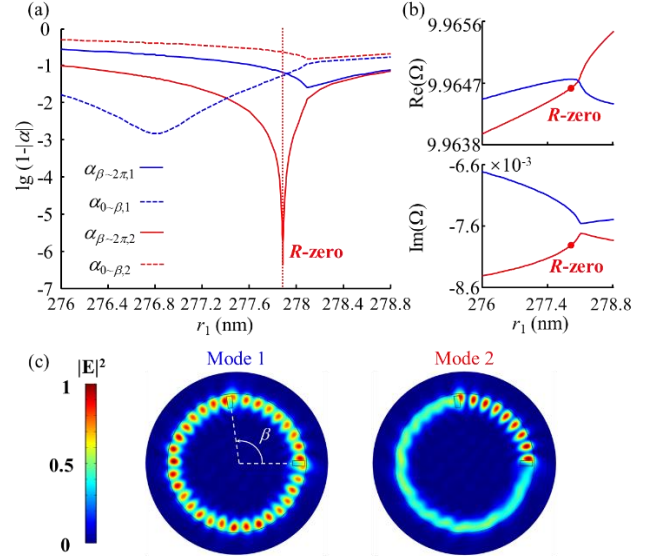

FIG. 2. (a) Local chiralities of the eigenmodes plotted as functions of the size $r_1$ of nanohole 1 around the EP. The solid and dashed curves represent the rigorous numerical results of $\alpha_{\beta\sim2\pi}$ and $\alpha_{0\sim\beta}$ (with an additional subscript 1 or 2 for eigenmode 1 or 2), respectively. (b) The dimensionless complex resonance frequencies $\Omega$ of the eigenmodes plotted as functions of $r_1$ around the EP. The blue and red curves in (a) and (b) represent the rigorous numerical results of the split eigenmodes 1 and 2, respectively. (c) The electric-field intensities of the pair of coalesced eigenmodes at the EP. The superimposed black lines show the boundaries of the cavity and nanoholes.

Here we provide the results of $\lg(1-|\alpha|)$ ($\alpha$ denoting $\alpha_{\beta\sim2\pi}$ or $\alpha_{0\sim\beta}$) to clearly show how close the local chirality is to $-1$ that corresponds to a perfect local chirality with a pure CCW APM. Thus, a smaller value of $\lg(1-|\alpha|)$ means that the local chirality is closer to be perfect.

Figure 2(a) shows that neither of the local chiralities $\alpha_{\beta\sim2\pi}$ and $\alpha_{0\sim\beta}$ is perfect at the EP. As $r_1$ changes, the maxima of $|\alpha_{\beta\sim2\pi}|$ and $|\alpha_{0\sim\beta}|$ locate away from the EP as indicated by the arrows shown in Fig. 2(a). An imperfect chirality indicates that both CW and CCW APMs exist in the eigenmodes at the EP. As shown in Fig. 2(c), slight interference fringes can be seen in the electric-field intensities of the pair of eigenmodes at the EP [31].

Next, we numerically solve the R-zeros $R_{cc}(\omega_R)=0$ at the effective scatterer and $R'_{cc}(\omega_R)=0$ at the complementary effective scatterer, respectively. By simultaneously scanning two parameters $r_1$ and $\beta$ around the two dips in Fig. 2(a), the resonant eigenmodes with locally perfect chirality $\alpha_{\beta\sim2\pi}\approx-1$ or $\alpha_{0\sim\beta}\approx-1$ are found away from the EP, which is achieved by making the complex APM coefficient $a_{cc}\approx0$ or $b_{cc}\approx0$ and corresponds to the R-zero $R_{cc}(\omega_R)=0$ or $R'_{cc}(\omega_R)=0$, respectively. A comparison between the structural parameters $r_1$ and $\beta$, dimensionless complex resonance frequency shifts $\Omega_{1(2)}-\Omega_0$ and local chiralities $\alpha_{\beta\sim2\pi,1(2)}$ and $\alpha_{0\sim\beta,1(2)}$ for the pair of eigenmodes at the EP and those at the R-zeros is provided in TABLE I, where $\Omega_0=9.8853162-0.0000603i$ is the dimensionless complex resonance frequency of the microcavity without the perturbation of nanoholes.

FIG. 3. (a) Local chiralities of the split eigenmodes plotted as functions of $r_1$ around the R-zero $R_{cc}(\omega_R)=0$ with $\alpha_{\beta\sim2\pi}\approx-1$. (b) The dimensionless complex resonance frequencies $\Omega$ of the split eigenmodes plotted as functions of $r_1$ around the R-zero. (c) The electric-field intensities of the pair of split eigenmodes with (right panel) and without (left panel) the locally perfect chirality $\alpha_{\beta\sim2\pi}\approx-1$.

Figures 3(a) shows the numerical results of local chiralities of the eigenmodes as functions of $r_1$ around the R-zero $R_{cc}(\omega_R)=0$ (indicated by the vertical red dotted line). It is seen that only one eigenmode (i.e., mode 2 shown by the red curves) exhibits a locally perfect chirality $\alpha_{\beta\sim2\pi}\approx-1$ (red solid curve) and $|\alpha_{0\sim\beta}|<1$ (red dashed curve) at the R-zero, while the other eigenmode (i.e., mode 1 shown by the blue curves) exhibits an imperfect chirality $|\alpha_{\beta\sim2\pi}|<1$ (blue solid curve) and $|\alpha_{0\sim\beta}|<1$ (blue dashed curve). The corresponding complex resonance frequencies of the pair of eigenmodes are plotted in Fig. 3(b). It is clearly seen that the complex resonance frequencies of the two eigenmodes split at the R-zero, implying that the system is away from the EP. The electric-field intensity of mode 2 with locally perfect chirality exhibits an ideal traveling wave pattern in the local azimuthal range $\beta\sim2\pi$, as shown in the right panel of Fig. 3(c), which is apparently different from that at the EP shown in Fig. 2(c). The electric-field intensity of mode 1 with imperfect chirality exhibits interference fringes in the global azimuthal range $0\sim2\pi$.

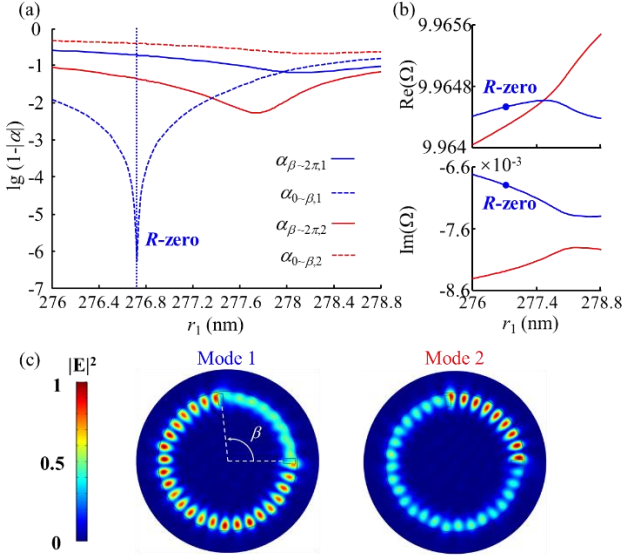

FIG. 4. (a)-(b) The same as Fig. 3 but for the $R$-zero $R'_{cc}(\omega_R) = 0$ with $\alpha_{0\sim\beta}\approx-1$. (c) The electric-field intensities of the pair of split eigenmodes with (left panel) and without (right panel) the locally perfect chirality $\alpha_{0\sim\beta}\approx-1$.

Figure 4 is similar to Fig. 3 but for the $R$-zero $R'_{cc}(\omega_R) = 0$. It is seen that only mode 1 shows a locally perfect chirality $\alpha_{0\sim\beta}\approx-1$ and $|\alpha_{\beta\sim2\pi}|<1$ at the $R$-zero. Correspondingly, the electric-field intensity of mode 1 shows an ideal travelling-wave pattern in the local azimuthal range $0\sim\beta$. Figures 3 and 4 indicate that the locally perfect chirality at the $R$-zero could exist in either the larger (Fig. 3) or the smaller (Fig. 4) azimuthal range divided by the two scatterers. This is quite different from the local chirality at the EPs, at which the stronger local chirality always exists in the larger azimuthal range [31].

For the model predictions, the $R$-zeros are obtained by solving two complex-valued Eqs. (8) taking $R_{cc}(\omega_R)=0$ or

TABLE II. Model predictions of the structural parameters $r_1$ and $\beta$, dimensionless complex resonance frequency shifts $\Omega_{1(2)}-\Omega_0$ and local chiralities $\alpha_{\beta\sim2\pi,1(2)}$ and $\alpha_{0\sim\beta,1(2)}$ at the EP (already obtained in Ref. [31]) and $R$-zeros.

|  | EP[31] | $R$-zero $R_{cc}(\omega_R)=0$ | $R$-zero $R'_{cc}(\omega_R)=0$ |
|---|---|---|---|
| $r_1$ (μm) | 0.2773424 | 0.2771580 | 0.2761291 |
| $\beta$ (rad) | 1.6821904 | 1.6821898 | 1.6821833 |
| $\Omega_1-\Omega_0$ | 0.0791424 | 0.0791858 | 0.0790612 |
|  | -0.0092223i | -0.008966i | -0.0085529i |
| $\Omega_2-\Omega_0$ | 0.0791419 | 0.0790038 | 0.0786209 |
|  | -0.0092246i | -0.0094729i | -0.0098114i |
| $\alpha_{\beta\sim2\pi,1}$ | -0.9831791 | -0.9370829 | -0.8080828 |
| $\alpha_{0\sim\beta,1}$ | -0.8665585 | -0.9428935 | *-0.9999999* |
| $\alpha_{\beta\sim2\pi,2}$ | -0.9834587 | *-0.9999999* | -0.9544870 |
| $\alpha_{0\sim\beta,2}$ | -0.8657942 | -0.7684105 | -0.5909489 |

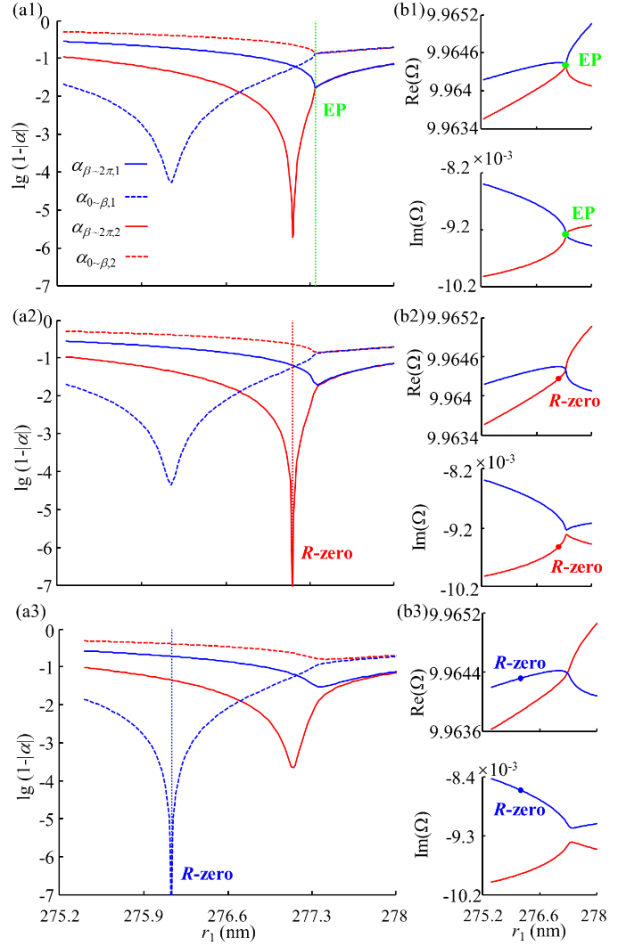

FIG. 5. Model predictions of local chiralities (a1) and the complex resonance frequencies (b1) of the pair of eigenmodes plotted as functions of $r_1$ around the EP. (a2)-(b2) and (a3)-(b3) are the same as (a1)-(b1) but for the $R$-zeros $R_{cc}(\omega_R) = 0$ and $R'_{cc}(\omega_R) = 0$, respectively.

Eqs. (15) taking $R'_{cc}(\omega_R) = 0$, which determine the unknowns of eigenfrequency $\omega_R$, $r_1$ and $\beta$. The EP is obtained by solving 3 complex-valued equations including Eq. (5) with $\omega$ replaced by $\omega_1$ or $\omega_2$ [corresponding to $\pm$ in Eq. (5), respectively] along with $\omega_1=\omega_2$, which determine the unknowns of eigenfrequency $\omega_1=\omega_2$, $r_1$ and $\beta$. The model predictions of the structural parameters, dimensionless complex resonance frequency shifts and local chiralities at the EP and the $R$-zeros are provided in TABLE II. Figures 5(a1)-(b1), (a2)-(b2) and (a3)-(b3) show the model predictions of local chiralities and complex resonance frequencies of the pair of eigenmodes as functions of $r_1$ around the EP, $R$-zeros $R_{cc}(\omega_R) = 0$ and $R'_{cc}(\omega_R) = 0$, respectively. It is seen that the model predictions agree well with the numerical results except slight deviations in predicting the structural parameters.

## B. Impact of nanohole size on the local chirality at the $R$-zeros

We now change the size of nanoholes to show the evolution of the local chiralities of the eigenmodes at the $R$-zeros. Here we consider the $R$-zero $R_{cc}(\omega_R)=0$ with locally perfect chirality $\alpha_{\beta\sim 2\pi}=-1$ for example. The size $r_2$ of nanohole 2 takes the values from $r_{2,0}=0.108816$ μm to $3r_{2,0}$ [31]. The numerical results of $R$-zeros are obtained by scanning $r_1$ and $\beta$ (near 130°) simultaneously.

As shown in Fig. 6(a), the numerical results of the local chiralities $\alpha_{\beta\sim 2\pi}$ (circles) and $\alpha_{0\sim\beta}$ (squares) of the eigenmodes at the $R$-zeros are plotted as functions of $r_2$. For the model predictions, the $R$-zeros are obtained by solving Eqs. (8). It is seen that the model predictions of $\alpha_{\beta\sim 2\pi}$ (solid curves) and $\alpha_{0\sim\beta}$ (dashed curves) agree well with the numerical results.

Figure 6(a) shows again that generally only one of the split eigenmodes has a locally perfect chirality $\alpha_{\beta\sim 2\pi}\approx-1$ (red circles and red solid curve) at the $R$-zero $R_{cc}(\omega_R)=0$. Accordingly, the electric-field intensity of the eigenmode with $\alpha_{\beta\sim 2\pi}\approx-1$ exhibits an ideal traveling-wave pattern within $\beta\sim 2\pi$ as shown in the second row in Fig. 6(b), which is not true for the other eigenmode as shown in the first row in Fig. 6(b).

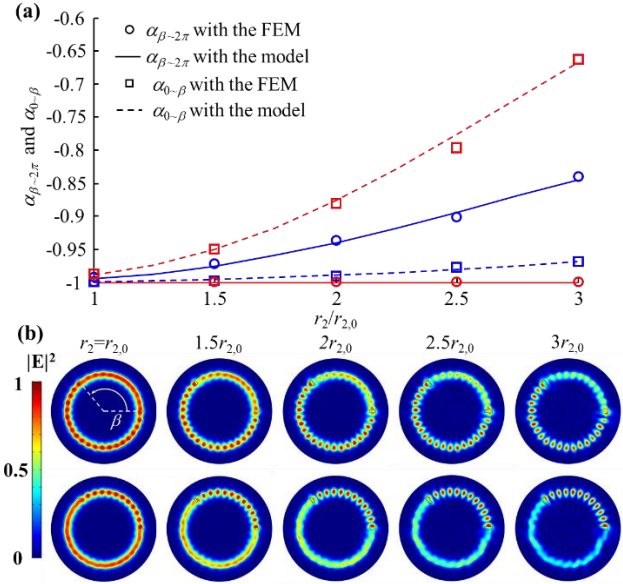

FIG. 6. (a) Local chiralities of the split eigenmodes at the $R$-zeros plotted as functions of the size $r_2$ of nanohole 2. Circles (squares) and solid (dashed) curves denote the numerical results and model predictions of $\alpha_{\beta\sim 2\pi}$ ($\alpha_{0\sim\beta}$), respectively. The blue and red curves correspond to two split eigenmodes at the $R$-zeros. (b) The electric-field intensities of the eigenmodes without (first row) or with (second row) the locally perfect chirality $\alpha_{\beta\sim 2\pi}\approx-1$ for the different $r_2$ shown in (a).

However, with the decrease of the nanohole size, both of the pair of eigenmodes tend to possess an approximately perfect global chirality over $[0, 2\pi]$, i.e., $\alpha_{0\sim\beta}\approx\alpha_{\beta\sim 2\pi}\approx-1$, implying an approximate coincidence of the $R$-zero $R'_{cc}(\omega_R)=0$ with $R_{cc}(\omega_R)=0$ for both split eigenmodes. As a result, the pair of split eigenmodes at the $R$-zeros approximately coalesce into one resonant eigenmode with an approximately perfect global chirality $\alpha_{0\sim\beta}\approx\alpha_{\beta\sim 2\pi}\approx-1$, which is simply the EP for weak scatterers [31]. As shown in Fig. 6(b), with the decease of the nanohole size, the interference fringes fade away and evolve into an ideal traveling-wave pattern over $0\sim 2\pi$ for both eigenmodes.

The approximate coincidence of the $R$-zero $R'_{cc}(\omega_R)=0$ with $R_{cc}(\omega_R)=0$ for weak scatterers can be explained by the model, i.e., by deriving $R'_{cc}=0$ from $R_{cc}=0$. By neglecting the higher-order multiple reflection at the (complementary) effective scatterer, the (complementary) effective scattering coefficients expressed as Eqs. (3) and (11) can reduce to

$$R_{cc}\approx\rho_2+v^2\rho_1\tau_2^2,\quad T\approx v\tau_1\tau_2, \quad (16a)$$
$$R'_{cc}\approx\rho_1+w^2\rho_2\tau_1^2,\quad T'\approx w\tau_1\tau_2. \quad (16b)$$

Substituting Eq. (16a) into $R_{cc}=0$ and $wT=1$ in Eqs. (8), one can obtain,

$$\rho_2+v^2\rho_1\tau_2^2\approx 0, \quad (17a)$$
$$wv\tau_1\tau_2\approx 1. \quad (17b)$$

Substituting Eq. (17b) into Eq. (17a) multiplied by $w^2\tau_1^2$, one can obtain $\rho_1+w^2\rho_2\tau_1^2\approx 0$, i.e., $R'_{cc}\approx 0$ according to Eq. (16b).

## C. Impact of relative azimuthal angle $\beta$ between the two nanoholes on the local chirality at the $R$-zeros

Next, we change the relative azimuthal angle $\beta$ between the two nanoholes with $r_2$ fixed to be $2.5r_{2,0}$ to show the evolution of the local chiralities of the split eigenmodes at the $R$-zero $R_{cc}(\omega_R)=0$. As shown in Fig. 7(a), the numerical results (hollow circles/squares) of the local chiralities of the eigenmodes at the $R$-zero plotted as functions of $\beta$ agree well with the model predictions (solid circles/squares). Figure 7(b) shows the electric-field intensities of 5 pairs of eigenmodes at the $R$-zeros with different $\beta$.

When $\beta>180°$, Fig. 7 shows again that the locally perfect chirality $\alpha_{\beta\sim 2\pi}\approx-1$ could occur in the smaller azimuthal range $\beta\sim 2\pi$ divided by the two scatterers for one of the split eigenmodes (red solid/hollow circles) at the $R$-zero, as already shown in Fig. 4. When $\beta$ is close to 180°, the other eigenmode with $|\alpha_{\beta\sim 2\pi}|<1$ tends to have a perfect chirality in $0\sim\beta$, i.e., $\alpha_{0\sim\beta}\approx-1$ (blue squares) as shown in the inset of Fig. 7(a) and for $\beta=173.340°$ in Fig. 7(b). This switch of the azimuthal range with perfect chirality between the split eigenmodes for $\beta$ close to 180° is consistent with the numerical results in Ref. [32].

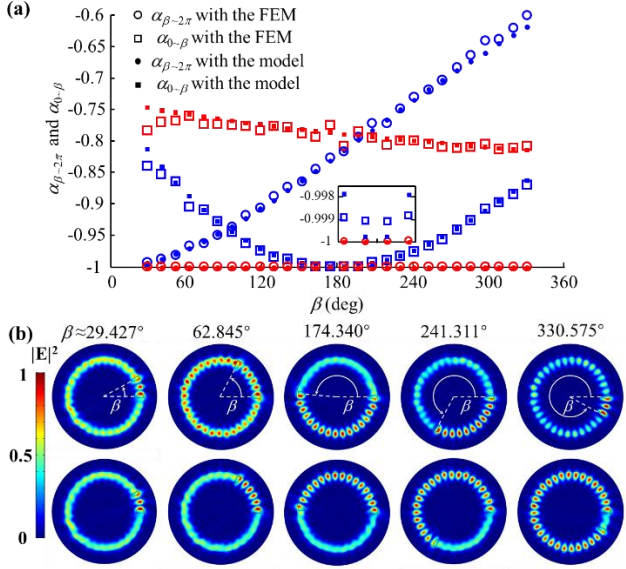

FIG. 7. (a) Local chiralities of the split eigenmodes at the $R$-zeros plotted as functions of the relative azimuthal angle $\beta$ between the two nanoholes. The hollow and solid cirlces (squares) represent the numerical results and model predictions of $\alpha_{\beta\sim 2\pi}$ ($\alpha_{0\sim\beta}$), respectively. The blue and red curves correspond to the two split eigenmodes at the $R$-zeros. The inset shows a magnification of the region $\beta \in [160°, 200°]$. (b) The electric-field intensities of the eigenmodes without (first row) or with (second row) the locally perfect chirality $\alpha_{\beta\sim 2\pi}\approx -1$ for the different $\beta$ shown in (a).

With the decease of $\beta$, the local chiralities of the pair of split eigenmodes at the $R$-zero tend to be identical. Consequently, the pair of split eigenmodes at the $R$-zero tend to coalesce into one eigenmode with an approximately perfect local chirality $\alpha_{\beta\sim 2\pi}\approx -1$, which is simply the EP under the condition that the two nanoholes have a small relative azimuthal angle [31]. Accordingly, the electric-field intensities of the pair of split eigenmodes tend to be identical and exhibit a traveling-wave pattern in $\beta\sim 2\pi$ as shown in Fig. 7(b).

Interestingly, Fig. 7(a) shows that for the eigenmode with locally perfect chirality $\alpha_{\beta\sim 2\pi}\approx -1$, the imperfect local chirality $\alpha_{0\sim\beta}$ takes a value from -0.8 to -0.75 (red hollow/solid squares) that weakly depends on $\beta$. This is further confirmed by the second row in Fig. 7(b), which shows that the electric-field intensities of the eigenmodes for different $\beta$ exhibit similar patterns in $0\sim\beta$ except for different number of interference nodes. This weak dependence of $\alpha_{0\sim\beta}$ on $\beta$ at the $R$-zero can be understood from the model equations. Substituting Eqs. (3) into Eqs. (8) [taking $R_{cc}(\omega_R)=0$], one can obtain,

$$\rho_2 + \frac{v^2 \rho_1 \tau_2^2}{1-v^2 \rho_1 \rho_2} = 0, \quad (18a)$$

$$w\frac{v\tau_1\tau_2}{1-v^2\rho_1\rho_2}=1. \quad (18b)$$

Substituting $v^2$ solved from Eq. (18a) into the square of Eq. (18b), one can obtain,

$$w^2 = \frac{\rho_1}{\rho_2}\left(\frac{\tau_2}{\tau_1}\right)^2 \frac{1}{\rho_2^2-\tau_2^2}. \quad (19)$$

Substituting Eq. (19) into the expression of $R'_{c(cc)}$ in Eq. (11) and also into Eq. (14), one can see that the local chirality $\alpha_{0\sim\beta}$ only depends on $\rho_j$ and $\tau_j$, which only depend on the parameters of nanoholes. Since $r_2$ is fixed and $r_1$ only changes slightly with $\beta$, one can conclude that $\alpha_{0\sim\beta}$ weakly depends on $\beta$.

## IV. CONCLUSION

In summary, locally perfect chiralities of resonant eigenmodes at the $R$-zeros away from EPs in a microcavity perturbed by two strong nanoscatterers of nanoholes are reported. With a first-principles-based model taking APMs as the local basis to describe the local chirality, the multiple scattering of the CW and CCW APMs at the nanoscatterers can be treated as a two-ports scattering problem at an effective or complementary effective scatterer under the cylindrical coordinate system, and the locally perfect chirality is predicted to result from the unidirectional reflectionlessness, i.e., $R$-zeros of the APMs at the effective or complementary effective scatterer.

Numerical results and model predictions consistently show that generally the $R$-zero locates away from the EP in the parameter space. Thus, the pair of split eigenmodes at the $R$-zero have different complex resonance frequencies and electromagnetic fields. Generally, only one of the split eigenmodes exhibits a locally perfect chirality at the $R$-zero. Different from the approximately perfect local chirality at the EP that only exists in the larger azimuthal range divided by the two scatterers [31], the exactly perfect local chirality at the $R$-zero can exist in either the larger or the smaller azimuthal range divided by the two scatterers, which holds for the $R$-zero at the effective or complementary effective scatterer, repectively. Interestingly, for the eigenmode with perfect local chirality at the $R$-zero, its imperfect local chirality in the other azimuthal range is found to weakly depend on the relative azimuthal angle $\beta$ between the two scatterers.

On the other hand, with the decrease of the nanoscatterer size or the relative azimuthal angle between the two nanoscatterers, the pair of split eigenmodes at the $R$-zero tend to coalesce into one eigenmode at the EP with an approximately perfect global or local chirality, respectively.

The present work provides a deep theoretical insight into the relationship among the local or global chirality, $R$-zeros and EPs. The perfect local chirality and frequency splitting at $R$-zeros may benefit potential applications of on-chip microcavity system with chiral eigenmodes, for instance, to realize unidirectional or bidirectional lasing at different cavity azimuthal positions or different frequencies.

ACKNOWLEDGMENTS

Financial support from the National Key Research and Development Program of China (No. 2022YFA1404602), National Natural Science Foundation of China (No. 12404438, No. 92250302, No. 62475120, No. 12404434), and 111 Project (No. B23045) is acknowledged.

Corresponding authors:
*liuht@nankai.edu.cn
†bofang@nankai.edu.cn
§taocan@nankai.edu.cn

[1] Y. Huang, Y. Shen, C. Min, S. Fan, and G. Veronis, Unidirectional reflectionless light propagation at exceptional points, Nanophotonics 6, 977 (2017).
[2] Z. Lin, H. Ramezani, T. Eichelkraut, T. Kottos, H. Cao, and D. N. Christodoulides, Unidirectional invisibility induced by PT-Symmetric periodic structures, Phys. Rev. Lett. 106, 213901 (2011).
[3] A. Regensburger, C. Bersch, M.-A. Miri, G. Onishchukov, D. N. Christodoulides, and U. Peschel, Parity-time synthetic photonic lattices, Nature (London) 488, 167 (2012).
[4] L. Feng, Y.-L. Xu, W. S. Fegadolli, M.-H. Lu, J. E. B. Oliveira, V. R. Almeida, Y.-F. Chen, and A. Scherer, Experimental demonstration of a unidirectional reflectionless parity-time metamaterial at optical frequencies, Nat. Mater. 12, 108 (2013).
[5] R. Colom, E. Mikheeva, K. Achouri, J. Zuniga-Perez, N. Bonod, O. J. F. Martin, S. Burger, and P. Genevet, Crossing of the branch cut: The topological origin of a universal 2π-phase retardation in non-Hermitian metasurfaces, Laser Photonics Rev. 17, 2200976 (2023).
[6] E. Mikheeva, R. Colom, K. Achouri, A. Overvig, F. Binkowski, J.-Y. Duboz, S. Cueff, S. Fan, S. Burger, A. Alù, and P. Genevet, Asymmetric phase modulation of light with parity-symmetry broken metasurfaces, Optica 10, 1287 (2023).
[7] H. Ramezani, H.-K. Li, Y. Wang, and X. Zhang, Unidirectional spectral singularities, Phys. Rev. Lett. 113, 263905 (2014).
[8] W. R. Sweeney, C. W. Hsu, S. Rotter, and A. D. Stone, Perfectly absorbing exceptional points and chiral absorbers, Phys. Rev. Lett. 122, 093901 (2019).
[9] C. Wang, W. R. Sweeney, A. D. Stone, and L. Yang, Coherent perfect absorption at an exceptional point, Science 373, 1261 (2021).
[10] J. Hou, J. Lin, J. Zhu, G. Zhao, Y. Chen, F. Zhang, Y. Zheng, X. Chen, Y. Cheng, L. Ge, and W. Wan, Self-induced transparency in a perfectly absorbing chiral second-harmonic generator, PhotoniX 3, 22 (2022).
[11] S. Soleymani, Q. Zhong, M. Mokim, S. Rotter, R. El-Ganainy, and S. K. Özdemir, Chiral and degenerate perfect absorption on exceptional surfaces, Nat. Commun. 13, 599 (2022).
[12] W. R. Sweeney, C.W. Hsu, and A. D. Stone, Theory of reflectionless scattering modes, Phys. Rev. A 102, 063511 (2020).
[13] X. Jiang, S. Yin, H. Li, J. Quan, H. Goh, M. Cotrufo, J. Kullig, J. Wiersig, and A. Alù, Coherent control of chaotic optical microcavity with reflectionless scattering modes, Nat. Phys. 20, 109 (2024).
[14] L. Chen and S. M. Anlage, Use of transmission and reflection complex time delays to reveal scattering matrix poles and zeros: Example of the ring graph, Phys. Rev. E 105, 054210 (2022).
[15] M. B. Soley, C. M. Bender, and A. D. Stone, Experimentally realizable PT phase transitions in reflectionless quantum scattering, Phys. Rev. Lett. 130, 250404 (2023).
[16] J. Sol, A. Alhulaymi, A. D. Stone, and P. Del Hougne, Reflectionless programmable signal routers, Sci. Adv. 9, eadf0323 (2023).
[17] Z. Rao, C. Meng, Y. Han, L. Zhu, K. Ding, and Z. An, Braiding reflectionless states in non-Hermitian magnonics, Nat. Phys. 20, 1904 (2024)
[18] F. Binkowski, F. Betz, R. Colom, P. Genevet, and S. Burger, Poles and zeros in non-Hermitian systems: Application to photonics, Phys. Rev. B 109, 045414 (2024).
[19] R. Huang, S. K. Özdemir, J.-Q. Liao, F. Minganti, L.-M. Kuang, F. Nori, and H. Jing, Exceptional photon blockade, Laser Photonics Rev. 16, 2100430 (2022).
[20] H. Lee, A. Kecebas, F. Wang, L. Chang, S. K. Özdemir, and T. Gu, Chiral exceptional point and coherent suppression of backscattering in silicon microring with low loss Mie scatterer, eLight 3, 20 (2023).
[21] J. N. Yang, S. S. Shi, S. Yan, R. Zhu, X. M. Zhao, Y. Qin, B. W. Fu, X. Q. Chen, H. C. Li, Z. C. Zuo, K. J. Jin, Q. H. Gong, and X. L. Xu, Non-orthogonal cavity modes near exceptional points in the far field, Commun. Phys. 7, 13 (2024).
[22] G. Chern, H. Tureci, A. D. Stone, R. Chang, M. Kneissl, and N. Johnson, Unidirectional lasing from InGaN multiple-quantum-well spiral-shaped micropillars, Appl. Phys. Lett. 83, 1710 (2003).
[23] J. Wiersig, Structure of whispering-gallery modes in optical microdisks perturbed by nanoparticles, Phys. Rev. A 84, 063828 (2011)
[24] B. Peng, S. K. Özdemir, M. Liertzer, W. J. Chen, J. Kramer, H.Yilmaz, J. Wiersig, S. Rotter, and L. Yang, Chiral modes and directional lasing at exceptional points, Proc. Natl. Acad. Sci. (USA) 113, 6845 (2016).
[25] Y. Chen, J. Li, K. Xu, S. Biasi, R. Franchi, C. Huang, J.

Duan, X. Wang, L. Pavesi, X. Xu, J. Wang, Electrically reconfigurable mode chirality in integrated microring resonators, Laser Photonics Rev. 18, 2301289 (2024).

[26] W. Mao, Z. Fu, Y. Li, F. Li, and L. Yang, Exceptional–point–enhanced phase sensing, Sci. Adv. 10, eadl5037 (2024).

[27] J. Zhu, Y. Zhong, and H. Liu, Impact of nanoparticle-induced scattering of an azimuthally propagating mode on the resonance of whispering gallery microcavities, Photon. Res. 5, 396 (2017).

[28] J. Zhu, H. Liu, F. Bo, C. Tao, G. Zhang, and J. Xu, Intuitive model of exceptional points in an optical whispering-gallery microcavity perturbed by nanoparticles, Phys. Rev. A 101, 053842 (2020).

[29] T. Kato, Perturbation Theory for Linear Operators (Springer-Verlag, Berlin, 1995).

[30] W. D. Heiss, The physics of exceptional points, J. Phys. A 45, 444016 (2012).

[31] J. Zhu, C. Wang, C. Tao, Z. Fu, H. Liu, F. Bo, L. Yang, G. Zhang, and J. Xu, Local chirality at exceptional points in optical whispering-gallery microcavities, Phys. Rev. A 108, L041501 (2023).

[32] J. N. Yang, H. C. Li, S. Yan, Q. H. Gong, and X. L. Xu, Local high chirality near exceptional points based on asymmetric backscattering, New J. Phys. 26, 093044 (2024).

[33] C. Vassallo, Optical Waveguide Concepts (Elsevier, 1991).

[34] H. Liu, DIF CODE for Modeling Light Diffraction in Nanostructures (Nankai University, Tianjin, 2010).